\documentclass[3p,times,procedia,letterpaper]{elsarticle} 

\usepackage{ecrc}

\usepackage{geometry}
\volume{00}

\firstpage{1}

\journalname{ISTTT Procedia}

\runauth{M.~Treiber, A.~Kesting}

\jid{sbspro}

\jnltitlelogo{Procedia Social and Behavioral Sciences}

\CopyrightLine{2016}{The Authors. Published by Elsevier Ltd. Selection and/or peer review under responsibility of Delft University of Technology}

\usepackage{amssymb}

\usepackage[figuresright]{rotating}

\usepackage{units} 

\providecommand{\bc}{\begin{center}}
\providecommand{\ec}{\end{center}}
\providecommand{\be}{\begin{equation}}
\providecommand{\ee}{\end{equation}}
\providecommand{\bea}{\begin{eqnarray}}
\providecommand{\eea}{\end{eqnarray}}
\providecommand{\bdm}{\begin{displaymath}}
\providecommand{\edm}{\end{displaymath}}
\providecommand{\bdma}{\begin{eqnarray*}}
\providecommand{\edma}{\end{eqnarray*}}
\providecommand{\ba}{\begin{eqnarray*}}
\providecommand{\ea}{\end{eqnarray*}}
\providecommand{\bi}{\begin{itemize}}
\providecommand{\ei}{\end{itemize}}
\providecommand{\benum}{\begin{enumerate}}
\providecommand{\eenum}{\end{enumerate}}

\providecommand{\refkl}[1]{(\ref{#1})}

\providecommand{\myVector}[1]{
  \left(\begin{array}{c} 
      #1 
   \end{array} \right)
}

\providecommand{\myMatrixTwo}[1]{
  \left(\begin{array}{cc} 
      #1 
   \end{array} \right)
}

\providecommand{\text}[1]{{\mbox{ #1}}}

\providecommand{\fig}[2]{
   \begin{center}
     \includegraphics[width=#1]{#2}
   \end{center}
}

\providecommand{\diff}[1]{ \ {\rm d} #1} 
\providecommand{\ablpart}[2]{\frac{\partial #1}{\partial #2}}  
\providecommand{\abl}[2]{\frac{{\rm d} #1}{{\rm d} #2}}  

\providecommand{\erw}[1]{\mbox{$\langle #1 \rangle$}}

\providecommand{\m}[1]{\text{\sf\textbf{#1}}}         

\providecommand{\sub}[1]{_{\rm #1}}
\renewcommand{\sup}[1]{^{\rm #1}}

\begin{document}

\begin{frontmatter}

\title{The Intelligent Driver Model with
    Stochasticity -- New Insights Into Traffic Flow Oscillations}

\author[TUD]{Martin Treiber\corref{cor1}}\ead{treiber@vwi.tu-dresden.de}
\author[TOMTOM,TUD]{Arne Kesting}\ead{mail@akesting.de}

\address[TUD]{Technische Universit\"at Dresden, Institute for Transport \& Economics,\\
W\"urzburger Str. 35, 01062 Dresden, Germany}
\cortext[cor1]{Corresponding author.}

\address[TOMTOM]{TomTom Development Germany,
An den Treptowers 1,
12435 Berlin (Germany)}

\begin{abstract}
Traffic flow oscillations, including traffic waves, are a common
yet incompletely understood feature of congested traffic. Possible
mechanisms include
traffic flow instabilities, indifference regions or finite human perception
thresholds (action points), and external acceleration
noise. However, the relative importance of these factors in a given
situation remains unclear. We bring light into this question by adding
external noise and action points to the
Intelligent Driver Model and other car-following models thereby obtaining a minimal model containing
all three oscillation mechanisms. We show analytically that even in the subcritical regime of linearly
stable flow (order parameter $\epsilon<0$), external white noise leads
to spatiotemporal speed correlations 
``anticipating'' the waves of the linearly unstable
regime. Sufficiently far away from the threshold, the amplitude scales
with $(-\epsilon)^{-0.5}$. By means of simulations and comparisons
with experimental car platoons and bicycle traffic, we show that 
external noise and indifference regions with 
action points have essentially equivalent effects. Furthermore,
flow instabilities dominate the oscillations
on freeways while external noise or action
points prevail at low desired speeds such as vehicular city or  
bicycle traffic. For bicycle traffic, noise can
lead to fully developed waves even for single-file traffic in
the subcritical regime.
\end{abstract}


\end{frontmatter}


\section{Introduction}
%
Traffic flow oscillations, including stop-and-go traffic, are a common
phenomenon in congested vehicular
traffic~\cite{Mauch-Cassidy,Ahn-Cassidy-OscillationsISTTT,Martin-empStates,Treiber-ThreePhasesTRB,Zielke-intlComparison}. 
Conventionally, this phenomenon is described in terms of linear or
nonlinear string or flow
instabilities~\cite{Wilson-Pattern2008,OroszPRE2009,TreiberKesting-book}
which are typically 
triggered by a local persistent 
perturbation, e.g., lane changes near a
bottleneck~\cite{Ahn-Cassidy-OscillationsISTTT}. In another approach,
the flow oscillations are traced back to indifference regions of the human
driver~\cite{Kerner2012physics,Jiang2014traffic}  or to finite perception thresholds
leading to abrupt acceleration changes at discrete ``action
points''~\cite{Wie74,Wagner-03}. Related to this are finite attention
spans~\cite{ThreeTimes-07,HDM}. It has also ben proposed that the
oscillations may be caused by event-oriented changes of the driving
style switching, e.g., between ``timid'' and
``aggressive''~\cite{laval2010mechanism}, or, related to this, by
over- and underreactions~\cite{yeo2009understanding}.  Finally, direct external
additive or multiplicative acceleration noise (e.g., caused by perception errors) is
postulated to drive the
oscillations. This line of reasoning 
is typically modelled by cellular automata
(e.g.,~\cite{TianTreiber2015}) which need some sort of 
stochasticity, anyway, for a proper specification. However, there are
also approaches to incorporate acceleration noise into time-continuous
car-following models leading to stochastic differential
equations~\cite{HDM,FPE-EuroPhys08}. One of the simplest approaches is
the ``Parsimonious Car-Following Model'' (PCF model)~\cite{laval2014parsimonious}
which adds white acceleration
noise to the free-acceleration part of Newell's car-following model
with bounded acceleration~\cite{laval2008microscopic} and,
as~\cite{HDM}, also
provides explicit numerical stochastic update rules by integrating the
stochastic 
differential equation over one time step. 

While all of the above approaches can explain certain
observations, it remains an open question whether these approaches are
connected with each other, and if so, in which way. Another open
problem is to identify the situations where
oscillations are caused predominantly by flow instabilities, by
indifference regions, or by noise. 

In this contribution, we bring light into these questions by proposing
a general scheme for adding
noise and indifference regions (in form of action points) to  a
class of deterministic acceleration-based car-following
models. Suitable underlying models include
the Intelligent Driver Model (IDM)~\cite{Opus}, the Full Velocity
Difference Model (FVDM)~\cite{Jiang-vDiff01}, or Newell's car-following
model~\cite{newell-carFollowing2002} with bounded accelerations~\cite{laval2008microscopic}. In this way, we obtain a minimal
model containing all three 
mechanisms which we then analyze analytically and numerically. The
focus is on the generic instability mechanisms and their relative importance
rather than on specific car-following models. 

The rest of the paper is organized as follows: In the next section,
we specify the minimal model. In Section~\ref{sec:analyt}, we
introduce the order parameter $\epsilon$ denoting the relative
distance to linear string instability and analytically derive, as a
function of $\epsilon$, the
statistical fluctuation properties induced by white noise, including
spectral, modal, and overall intensity of the vehicle gap and speed
fluctuations, and the associated spatiotemporal correlations. In the Sections~\ref{sec:platoon}
and~\ref{sec:bicycle}, we investigate the oscillation mechanisms for
high-speed  and low-speed traffic (cars and bicycles, respectively),
and compare the results with experimental observations. Finally,
Section~\ref{sec:concl} concludes with a discussion.

\section{\label{sec:meth}Model Specification}
%
We consider general stochastic time-continuous car-following models of
the form
\be
\label{micGen}
\dot{v}_n=f(s_n,v_n,v_l)+\xi_n(t), \quad
\erw{\xi_n(t)}=0, \quad
\erw{\xi_n(t)\xi_m(t')}=Q\delta_{nm}\delta(t-t').
\ee
Here, $f(\cdot)$ denotes the  acceleration function of the underlying
car-following model for
vehicle $n$ as a
function of the (bumper-to-bumper) gap $s_n$, the speed $v_n$, and the
leader's speed $v_l$. Time delays in the independent variables representing reaction times such as
in Newell's Car-Following Model~\cite{newell-carFollowing2002} are allowed.
The white acceleration noise $\xi_n(t)$ is
completely uncorrelated in time and between vehicles (the Kronecker symbol
$\delta_{nm}=1$ for $n=m$ and zero, otherwise; $\delta(t-t')$ denotes
Dirac's delta distribution), and has the intensity
$Q$. Model~\refkl{micGen} can be seen as a simplistic special case of
the Human Driver Model (HDM)~\cite{HDM}. Notice that, when starting
with a deterministic initial state $v_n(0)$ at 
$t=0$, integration of the stochastic differential
equation~\refkl{micGen} leads, in the limit $t\to
0$, to a Gaussian speed distribution whose expectation and variance
are given by (see, e.g., \cite{honerkampEngl})
\be
\label{erwVv}
\erw{v_n(t)}=v_n(0)+f t, \quad
\erw{(v_n(t)-(v_n(0)+f t)^2}=Qt
\ee
where the variance is independent of $f$.
 Consequently, $Q$ has the unit $\unit{m^2/s^3}$.  
By chosing a deterministic car-following model with the ability for
string instability, e.g., the Intelligent Driver
Model (IDM)~\cite{Opus} or the Full Velocity Difference Model
(FVDM)~\cite{Jiang-vDiff01}, the model~\refkl{micGen} contains the two
oscillation-inducing factors noise and string instability. We
introduce the third factor, indifference regions in the form of
action points, by updating the 
deterministic acceleration to the actual value given by $f(\cdot)$ only, if 
\be
\label{IDMaction}
\left|f(s_i(t),v_i(t),v_l(t)) - f(s_i(t'),v_i(t'),v_l(t'))\right|>\Delta a, \quad
\Delta a \sim U(0, \Delta a\sub{max}).
\ee
Here, $t'$ is the time of the last change (action point) and $\Delta
a$ is a uniformly distributed random number drawn at the last action
point whose maximum value must be well below the maximum acceleration
capability of the deterministic model. 

\begin{figure}
\fig{0.55\textwidth}{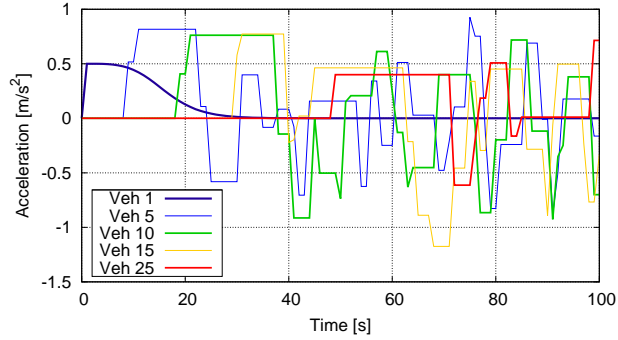}
\caption{\label{fig:actPoints}Visualization of the action points
  (maximum step $\Delta a\sub{max}=\unit[1.0]{m/s^2}$) by a
  time series of the accelerations of four platoon vehicles
  corresponding to Fig.~\ref{fig:IDMplatoon}(e) and (f). Also shown is the
  leader (Vehicle~1) accelerating slowly and 
  deterministically to the final speed of \unit[30]{km/h}.
   }
\end{figure}

Since characterizing action points according
to model~\refkl{IDMaction} is a novel proposition in itself, we first show
how the mechanism works by displaying typical instances of the
resulting acceleration time series
(Fig.~\ref{fig:actPoints}).  As expected, the drivers drive at
constant accelerations, most of the time. These episodes with typical
irregular durations between \unit[1]{s} and \unit[20]{s} are separated
by ``action points''  where the
acceleration is changed by a variable amount whose maximum is given
by the parameter $\Delta a\sub{max}$. Notice that both the irregular
durations and increments agree with observations~\cite{Wagner-03}.

The Eqs.~\refkl{micGen} (or~\refkl{erwVv}), \refkl{IDMaction}, and
 a deterministic car-following model specify the proposed minimal  
model containing all three oscillation factors. Each factor can be
controlled by a single parameter. The noise is controlled by the noise
intensity $Q$ (typical values are of the order of
$\unit[0.2]{m^2/s^3}$), the indifference region by the maximum
acceleration step $\Delta a\sub{max}$ (of the order of
$\unit[1]{m/s^2}$ or less), and the string instability by the relevant
parameter of the underlying car-following model, e.g., the maximum
acceleration $a$ for the IDM. It is
convenient to define the relative distance to the linear threshold
$a_c$ (see Section~\ref{sec:analyt} below) by
the order parameter
\be
\label{eps}
\epsilon=1-\frac{a}{a_c}
\ee
keeping the other IDM parameters fixed.
A homogeneous deterministic steady state is linearly stable for
$\epsilon<0$ (subcritical regime), and linearly string unstable for
$\epsilon>0$ (supercritical).

\section{\label{sec:analyt}Noise-Induced Subcritical Oscillations}
%
In order to analytically determine the linear response to the white
acceleration noise, we assume a homogeneous ring road of
circumference $L$ with $N$ identical
drivers/vehicles such that the global density is given by
$\rho=N/L$. Furthermore, we
switch off the action points by setting $\Delta 
a\sub{max}=0$.  We generalize the standard linear stability analysis
(see, e.g., Ref.~\cite{TreiberKesting-book}) to include noise. Starting
from a homogeneous steady state $v_n=v_e$ and $s_n=s_e(v_e)$
lying on the fundamental diagram, we decompose the gaps and speeds into
the steady-state contribution and a small time-dependent perturbation by setting $s_n(t)=s_e+y_n(t)$,
$v_n(t)=v_e+u_n(t)$. This leads to the linearized equations
\be
\label{lin}
\abl{y_n}{t} = u_{n-1}-u_n, \quad 
\abl{u_n}{t} = f_s y_n+f_v u_n+f_l u_{n-1}+\xi_n(t),
\ee
where  $f_s=\ablpart{f}{s}$, $f_v=\ablpart{f}{v}$, and
$f_l=\ablpart{f}{v_l}$ are the gradients of the acceleration function
$f(\cdot)$ at the deterministic steady state.

The time-dependent parts can be further decomposed into $N$ harmonic
\textit{eigenmodes} of wave number $k=2\pi m/N$, $m=-N/2+1, ..., N/2$
(with $|m|$ indicating the number of travelling waves on the ring and
$2\pi/|k|$ the number of vehicles per wave) 
 by the \textit{ansatz}
\be
\myVector{y_n(t)\\u_n(t)}=\sum_k \myVector{\hat{y}_k(t)\\ \hat{u}_k(t)} 
e^{ink}
\ee
where $i=\sqrt{-1}$ is the imaginary unit.
Inserting this into~\refkl{lin}, we obtain a set of independent 
pairs of stochastic differential equations for the modes $k$, 
\be
\label{lin-k}
\begin{array}{rcl}
\abl{\hat{y}_k}{t} &=& \left( e^{-ik}-1\right)\hat{u}_k, \\
\abl{\hat{u}_k}{t} &=& f_s \hat{y}_k+\left(f_v+f_l e^{-ik}\right)\hat{u}_k
+ \hat{\xi}_k(t),
\end{array}
\ee
where the noise source now is given by
\be
\label{noise-k}
\left\langle \hat{\xi}_k(t) \right\rangle=0, \quad
\left\langle \hat{\xi}_k(t)\hat{\xi}_l(t') \right\rangle
=\frac{Q}{N}\delta_{kl}\delta(t-t').
\ee

\begin{figure}
\fig{0.99\textwidth}{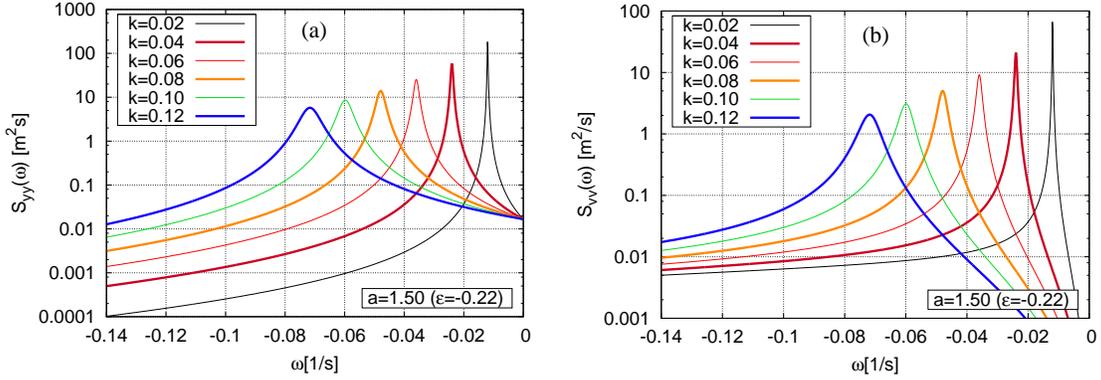}
\caption{\label{fig:analyt_eps1}Steady-state fluctuation spectrum of
  (a) the speeds, and~(b) the gaps for the IDM with the parameters of the main text and white acceleration
  noise of intensity 
  $Q=\unit[0.1]{m^2/s^3}$. Shown is the spectrum for several modes of
  dimensionless wavenumber $k$ (number of vehicles $2\pi/k$ per wave).
  The negative values for $\omega$ indicate an upstream propagation of
  $-\omega/k$ vehicles per second in the frame comoving with the
  vehicles. The underlying deterministic steady state is
  characterized by $v_e=\unit[48]{km/h}$ corresponding to a density
  of \unit[36.4]{vehicles/km}.
 }
\end{figure}

Each pair of linear stochastic differential
equation~\refkl{noise-k} can be written in the general form
\be
\label{linstoch}
\abl{}{t}\vec{X}_k = -\m{L}_k)\vec{X}_k + \vec{\xi}, \quad
\erw{\vec{\xi}(t)}=0, \quad
\erw{\vec{\xi}(t)\vec{\xi}'(t')}=\m{D}\delta(t-t')
\ee
where $\vec{X}_k=(\hat{y}_k, \hat{u}_k)'$ denotes the state vector of
mode $k$, and
\be
\label{LD}
\m{L}_k=\myMatrixTwo{0 & 1-e^{-ik}\\-f_s & -(f_v+f_le^{-ik})}, \quad
\m{D}=\frac{Q}{N}\myMatrixTwo{0&0\\0&1},
\ee
denote the linear dissipation
matrix, and the covariance matrix of the 
white noise, respectively.

The stationary solutions to~\refkl{linstoch} consist of
zero-mean Gaussian
fluctuations which are therefore completely specified by the amplitude
and correlation of the state variables as well as the temporal
autocorrelation function. 
Instead of the autocorrelation function, one
can also determine the spectral intensity of the gap and speed
oscillations of waves of wavenumber $k=2\pi/L$ (where $L$ is the
wavelength), i.e., the differential energy content (amplitude 
squared) contained in waves of wavenumber $k$ at an
angular frequency $\omega=2\pi/T$ (where $T$ is the wave period).
Remarkably, there exists an analytic
solution for the spectral intensity of the
stationary  fluctuations called the fluctuation-dissipation
theorem~\cite{honerkampEngl},
\be
\label{fluctdiss}
\m{S}_k(\omega)=\left(\m{L}_k+i\omega\m{1}\right)^{-1} \m{D}
\left(\m{L}'_k-i\omega\m{1}\right)^{-1}.
\ee
Here,
$\m{S}_k(\omega)=\tilde{\vec{X}}_k(\omega)\tilde{\vec{X}}_k'(\omega)$ is
the spectral intensity with $\tilde{\vec{X}}_k(\omega)$ the Fourier
transform of $\vec{X}_k(t)$, and $\tilde{\vec{X}}_k'(\omega)$ its transposed
complex conjugate. Furthermore, $\m{L}_k'$ is the transposed and complex conjugate of the linear
matrix $\m{L}_k$. Inserting~\refkl{LD}, we finally obtain the
subcritical modal fluctuation spectrum
\be
\label{S}
\m{S}_k(\omega)=\frac{Q}{N|\text{Det}(\m{L}+i\omega\m{1})|^2}
\myMatrixTwo{2(1-\cos k) & i\omega(1-e^{-ik})\\
-i\omega(1-e^{ik}) & \omega^2}
\ee
where
\bdm
\left|\text{Det}(\m{L}+i\omega\m{1})\right|^2=
 \left(\omega^2+\omega f_l \sin k-f_s(1-\cos k)\right)^2
+ \left( f_s \sin k - \omega(f_v+f_l \cos k)\right)^2.
\edm
The diagonal components are of particular
interest. $S_{11}=S_{yyk}(\omega)$ gives the spectral intensity of the
gap oscillations contained in waves of wavelength $2\pi/(\rho |k|)$
and period $2\pi/\omega$ propagating at a velocity $\omega/(\rho k)$
in the comoving frame of reference. The component
$S_{22}=S_{uuk}(\omega)$ gives the corresponding spectral intensity of the
speed fluctuations. Since the inverse Fourier transform of the
spectral intensity with respect to $\omega$ and $k$ gives the 
spatiotemporal correlation function, the fluctuation spectrums
$S_{yyk}(\omega)$ and 
$S_{uuk}(\omega)$ completely describe the subcritical linear
gap and speed fluctuations of the vehicles.

Figure~\ref{fig:analyt_eps1} displays 
$S_{yyk}(\omega)$ and $S_{uuk}(\omega)$ for the IDM parameters
$v_0=\unit[30]{m/s}$, $T=\unit[1.5]{s}$, $s_0=\unit[2]{m}$,
$b=\unit[1.5]{m/s^2}$ and a vehicle length of \unit[5]{m} at the
steady state $v_e=\unit[48]{km/h}$ corresponding to a density
  $\rho=\unit[36.4]{veh./km}$. For this state, the linear stability condition 
  $2v'_e(s_e)<f_l-f_v$ (cf.~\cite{TreiberKesting-book}) leads to the
  critical IDM acceleration parameter 
  $a_c=\unit[1.233]{m/s^2}$. Consequently, the acceleration
  $a=\unit[1.5]{m/s^2}$ of these plots corresponds to the control
  parameter $\epsilon=1-a/a_c=-0.22$. Remarkably, the spectrum of both
  the gap and the speed modes  shows distinct peaks
  at approximatively $\omega/k=\unit[-0.6]{s^{-1}}$. Since $2\pi/k$
  denotes the number of vehicles in a wave, $\omega/k$ can be
  interpreted as the number of vehicles per time unit a wave passes (``passing
  rate''). Thus, the fluctuations are highly spatio-temporally
  correlated propagating backwards at 0.6 vehicles per second, even
  significantly below the linear threshold. Notice that fully
  developed supercritical
  fluctuations triggered by linear instabilities have essentially the
  same passing rate.

\begin{figure}
\fig{0.99\textwidth}{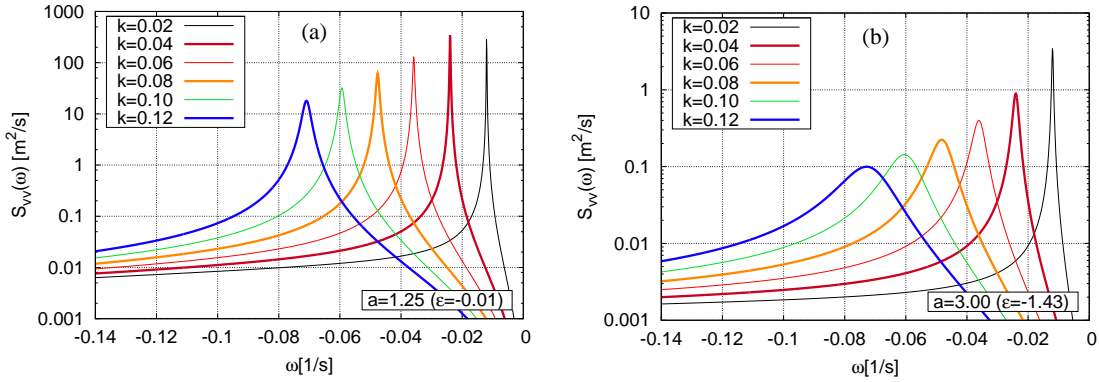}
\caption{\label{fig:analyt_eps03}Steady-state speed fluctuation
  spectrum for the control parameters 
  $\epsilon=0.01$~(a) and  
  $\epsilon=1.43$~(b). The IDM parameters, noise intensity, and the
  deterministic steady state is the same as in
  Figure~\ref{fig:analyt_eps1}.
}
\end{figure}

\begin{figure}
\fig{0.99\textwidth}{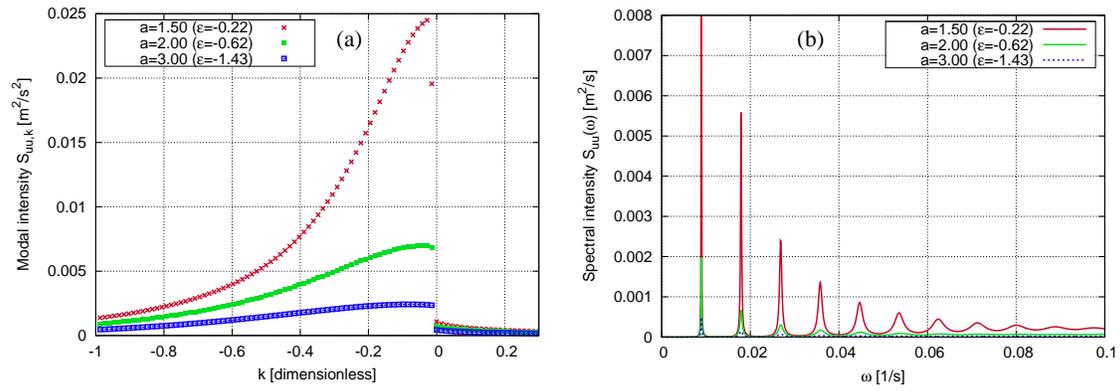}
\caption{\label{fig:analyt_integr_spectr}(a)~Modal speed fluctuation
  intensity for a ring road of length $L\sub{ring}=\unit[10]{km}$ for three
  values of the relative proximity $\epsilon$ to the linear stability
  threshold. Each symbol denotes a possible mode $k=2n\pi/L\sub{ring}$
  (negative if propagating backwards). (b)~Overall spectral
  intensity~\refkl{SuuOm}. The IDM parameters, noise intensity, and the
  deterministic steady state are the same as in
  Figure~\ref{fig:analyt_eps1}.
 }
\end{figure}

\begin{figure}
\fig{0.6\textwidth}{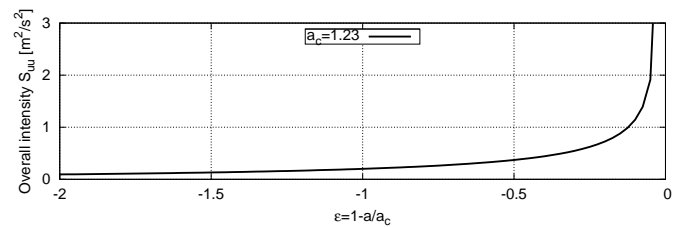}
\caption{\label{fig:analyt_intens}Overall speed fluctuation
  intensity~\refkl{Suu} as a function of the control parameter
  $\epsilon$ for the system of Figure~\ref{fig:analyt_integr_spectr}.
}
\end{figure}

Figure~\ref{fig:analyt_eps03} confirms this observations for other
order parameters. While it is expected that the spectral peaks
become more pronounced very near the threshold ($\epsilon=-0.01$),
significant peaks remain for $\epsilon=-1.43$, i.e., deep in the
stable regime.

Figure~\ref{fig:analyt_integr_spectr} displays one-dimensional
integrations of the modal speed fluctuation spectrum for a ring road
of length $L\sub{ring}=\unit[10\,000]{m}$. The panel~(a) shows the
modal fluctuation intensity 
\be
\label{Suuk}
S_{uuk}=\int S_{uuk}(\omega) \diff{\omega}
\ee
for some of the allowed modes near $k=0$. We observe that the
integration eliminates the nontrivial peaks which is to be expected:
After all, in the subcritical regime, the linear relaxation rate
decreases with decreasing absolute values of the wavenumber thereby
increasing, according to the 
fluctuation-dissipation theorem, the fluctuation intensities. Notice,
however, that a distinct anisotropy favoring backwards propagating
waves remain.  

Panel~\ref{fig:analyt_integr_spectr}(b) displays  the overall spectral intensity
\be
\label{SuuOm}
S_{uu}(\omega)=\sum_k S_{uuk}(\omega)
\ee
where the sum includes all allowed modes of the closed system. In
contrast to plot\ref{fig:analyt_integr_spectr}(a), distinct spectral peaks remain indicating
resonances of the lowest allowed modes. The distance between the peaks
is inversely proportional to the ring circumference.

\begin{figure}
\fig{0.8\textwidth}{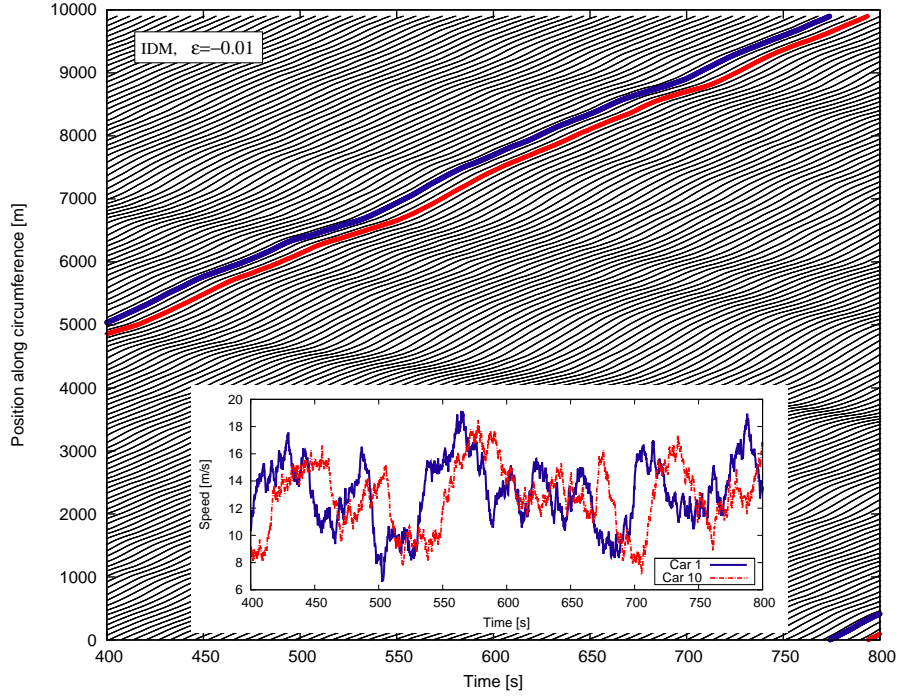}
\caption{\label{fig:IDMstochRing}Simulated IDM trajectories for
 a ring of \unit[10]{km} circumference, $a=\unit[1.25]{m/s^2}$
 ($\epsilon=-0.01$), and the remaining IDM parameters as in the main
 text. Every 3rd vehicle is shown. The inset shows speed time series
 of two selected vehicles. 
   }
\end{figure}

Finally, Figure~\ref{fig:analyt_intens} displays the overall
steady-state fluctuation
  intensity
\be
\label{Suu}
S_{uu}=\sum_k \int S_{uuk}(\omega) \diff{\omega}
\ee
of the speed of every single vehicle 
which is equal to the speed variance.
 Sufficiently far
away from the threshold ($\epsilon <-0.2$), the variance is inversely
proportional to $(-\epsilon)^{-1}$ corresponding to amplitudes
proportional to $(-\epsilon)^{-0.5}$. Notice that the variance does
not diverge at the linear stability limit which is caused (i) by
finite-size effects, (ii) by nonlinear saturation. As expected, the speed
fluctuation amplitude of a given vehicle is essentially
independent of the system size.

In summary, the analytical investigation shows that the 
fluctuations triggered by white noise
in the subcritical regime ``anticipate'' the characteristics of
traffic waves in the collectively unstable regime. This is examplified in
the simulated trajectories of Fig.~\ref{fig:IDMstochRing} whose
amplitude and spatiotemporal correlations agree quantitatively with
the analytic theory.  Moreover, the
amplitude of the subcritical fluctuations increases strongly when
approaching 
the linear threshold from below. The simulations in the following
sections demonstrate that this can even lead to fully developed waves
in the subcritical regime.

\section{\label{sec:platoon}Vehicular Traffic: Platoon Experiments}
%
In this section, we simulate platoon car-following experiments and
test which of the three possible oscillation mechanisms, namely instability, noise, or action
points, or  which combinations thereof, allows for reproducing
the empirical findings of a concave increase of the speed fluctuation
amplitude as a function of the platoon vehicle
number~\cite{tian2016empirical}). 
Moreover, by simulating these mechanisms with three underlying models
(the IDM, the FVDM and the PCF model),
we test to which extent the results are universal, i.e., independent
of the specific car-following model. In the IDM and FVDM simulations,
the white 
acceleration noise is integrated according to~\refkl{erwVv} resulting
in fluctuations that are asymptotically 
independent of the simulation update time step~\cite{HDM}. The PCF
model is simulated according to~\cite{laval2014parsimonious}, i.e.,
realisations of the analytical distributions of the
displacements are added to the locations in each time step which is
set equal to its time-gap parameter $\tau=T=\unit[1]{s}$ (see below), so no
discretisation errors incur. 
 
\begin{figure}
\fig{0.99\textwidth}{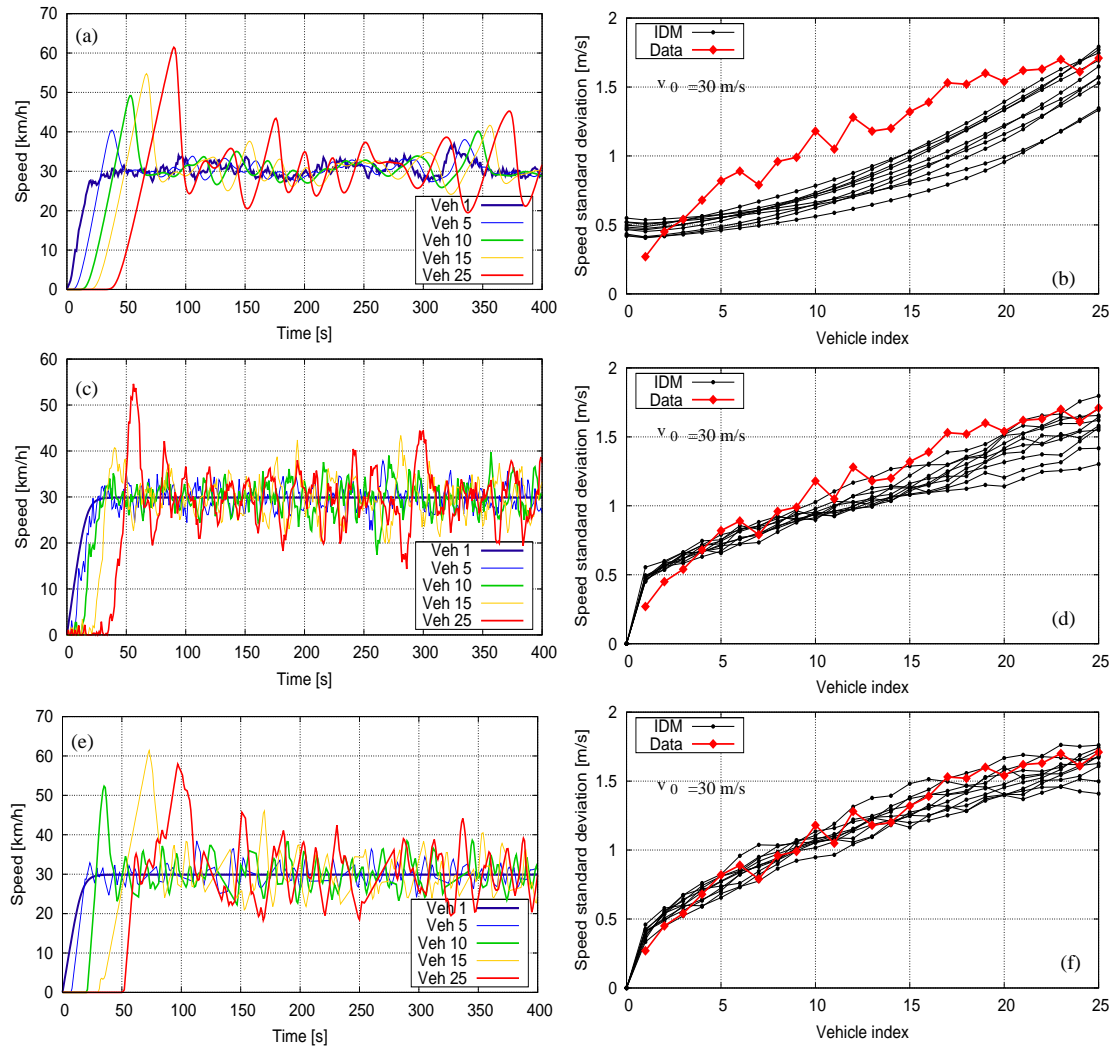}
\caption{\label{fig:IDMplatoon}Simulated speed time series (left column) and
  speed standard deviations (right column) for
  several vehicles in a platoon following the leader (Vehicle~1)
  accelerating slowly to the leading speed of \unit[30]{km/h}. The
  simulation setup reproduces the platoon experiments
  of~\cite{Tian2016560} (red symbols at the right column). In order
  to eliminate initial transients, the first \unit[200]{s} of the
  simulation are skipped when calculating standard deviations.
  Top row:
  deterministic IDM far in the unstable regime ($a=\unit[0.5]{m/s^2}$
  corresponding to $\epsilon=0.6$; some noise has been added to the
  leader to initiate oscillations); middle row: strong acceleration
  noise $Q=\unit[0.32]{m^2/s^3}$ at marginal stability ($\epsilon=0$) and
  without action points ($\Delta a\sub{max}=0$); bottom row: only
  action points ($\Delta a\sub{max}=\unit[1.2]{m/s^2}$, $Q=0$) at
  marginal stability. The remaining IDM parameters of the 
  followers are given in the main text.
}
\end{figure}

Figure~\ref{fig:IDMplatoon} displays simulations of a platoon car-following
experiment in China~\cite{Tian2016560} (red symbols in the right
column) where, in each row, only one of the three instability
mechanisms is activated, namely string instability (top row), white
acceleration 
noise (middle), and action points (bottom). 
Because of the stochastic
nature, the detailed dynamics changes from run to run, so we show the growth
of the speed standard deviation along the platoon  for 10
realisations (right column).  
Additionally, we show in the left column the time series of a typical run.  
The leader drives fully
deterministically (middle, bottom row) or with white noise (top row) according
to the IDM and freely accelerates from speed zero at the begin of the
experiments to a desired speed of $v_0\sup{lead}=\unit[30]{km/h}$ corresponding
to the experiment. The followers have a significantly higher desired speed and
initially stand in a queue behind the leader. The IDM
parameters of the  
  followers that are not directly related to the oscillation mechanisms are
  set always to $v_0=\unit[108]{km/h}$, $T=\unit[1.0]{s}$,
  $s_0=\unit[2]{m}$, and $b=\unit[2]{m/s^2}$ (notice that the vehicle
  length is 
  irrelevant in platoon simulations). The three oscillation
  mechanisms are controlled by the relative IDM acceleration 
$\epsilon=1-a/a_c$ (where
  $a_c=\unit[1.25]{m/s^2}$ for the above parameters and a leading
  speed of \unit[30]{km/h}),
  $Q$ (noise intensity), and $\Delta a\sub{max}$ (maximum acceleration
  change at an action point). In the simulations of each of the three
  mechanisms, the respective control parameter $\epsilon$, $Q$, and
  $\Delta a\sub{max}$ 
  has been calibrated to minimize the SSE between the observed and
  simulated speed standard deviations of all the vehicles (right
  column) over 10 simulation
  runs with independent seeds while the other two control parameters
  have been set to zero. 

We find that the 
instability mechanism alone (Panels (a) and (b)) leads to
oscillations that do not agree qualitatively with the
platoon experiments. Notably, the increase of the fluctuation
amplitude along the platoon vehicles is convex instead of
concave. Furthermore, the calibrated IDM acceleration parameter
$a=\unit[0.5]{m/s^2}$, or $\epsilon=0.6$, corresponds to an unrealistically
unstable regime which leads to unrealistic results in simulation
runs with longer platoons. Finally, the result depends sensitively on
the noise intensity of the leading vehicle with no sensible results
(no fluctuations)
obtained for zero noise.

Panels~\ref{fig:IDMplatoon}(c) and~(d) display the effect of external
noise for a marginal stability $\epsilon=0$ and no action points, $\Delta a\sub{max}=0$. In agreement
with observations, we obtain a concave growth of the fluctuation
amplitude along the platoon vehicles which, for
$Q=\unit[0.32]{m^2/s^3}$,  nearly quantitatively
agrees with the experiment. Notice that the instability mechanism plays a
role as well: The best results are obtained near marginal stability. 
Panels~\ref{fig:IDMplatoon}(e) and~(f) give the results of simulations with active action
points ($\Delta a\sub{max}=\unit[1.2]{m/s^2}$) and deactivated
noise ($Q=0$). Again, simulations near marginal stability ($\epsilon=0$)
give the best results. Compared  to
the simulations with pure acceleration noise, the fit quality is even
better. For most realisations,  a good agreement is reached for the first
platoon vehicles as well while the IDM with acceleration noise systematically
overestimates the speed standard deviation of these vehicles.

We conclude that the instability mechanism alone is
not able to reproduce the observations, even in the presence of  a fluctuating leader.
In contrast, both action points in the form of model~\refkl{IDMaction}
and white noise can reproduce the observations with best results
obtained near marginal stability. Remarkably, action points
and white noise are essentially interchangeable with action points
giving marginally better results.

\begin{figure}
\fig{0.99\textwidth}{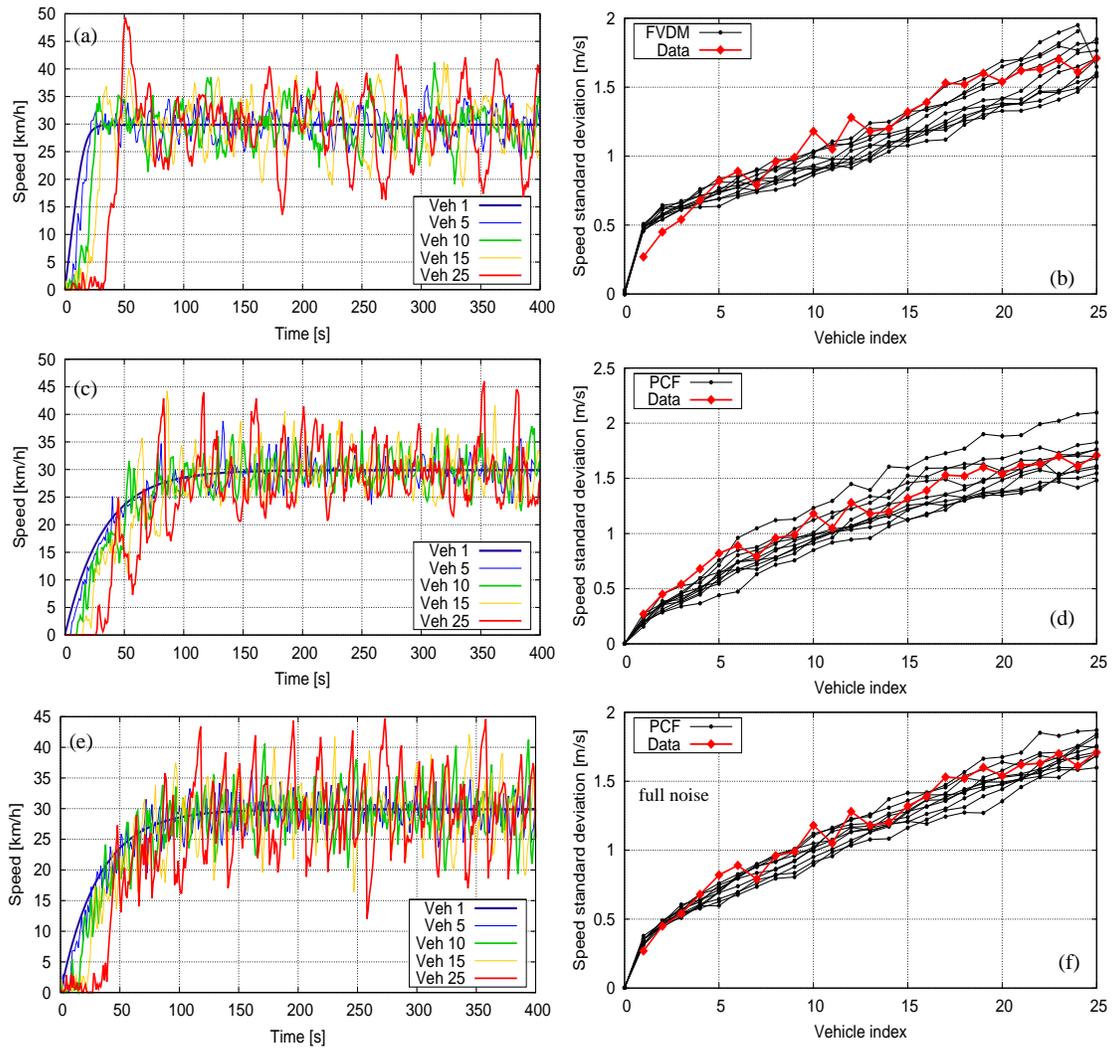}
\caption{\label{fig:otherplatoon}Speed time series (left column) and
  speed standard deviations (right column) for
  the same initial configuration and the same leader's speed profile
  as in Fig.~\ref{fig:IDMplatoon}.
  Top row: stochastic Full Velocity Difference Model (SFVDM) with a triangular
  fundamental diagram  for calibrated
  model parameters; middle row: Parsimonious Car-Following 
  Model (PCF model)~\cite{laval2014parsimonious} with the same
  fundamental diagram as the SFVDM. Bottom row: PCF model with
  the same parameters but stochasticity also turned on in the interacting
  regime. The values of the model
  parameters are given in the main text.
}
\end{figure}

The question arises if these results depend on the underlying model,
namely the IDM, or if other, possibly simpler, car-following models
can be used as well. In order to test this proposition, we have
performed simulations with the stochastic Full Velocity Difference
Model (SFVDM) and also with the PCF model~\cite{laval2014parsimonious} of
Laval et al which essentially implements the acceleration noise
mechanism at marginal stability without interaction points. 
 Instead of the optimal velocity (OV) function of the
original FVDM~\cite{Jiang-vDiff01}, we apply an OV function
corresponding to a tridiagonal fundamental diagram which resembles the
IDM and has the same desired speed, 
desired time gap and minimum gap parameters. 
The resulting
acceleration function of the FVDM reads
\be
\label{FVDM}
f(s,v,v_l)=\beta\big(v\sub{opt}(s)-v\big)+\lambda(v_l-v), \quad
v\sub{opt}(s)=\max \left(0, \min \left(v_0, \frac{s-s_0}{T}\right)\right)
\ee
with values of the steady-state parameters 
$v_0=\unit[30]{m/s}$, $T=\unit[1.0]{s}$, and $s_0=\unit[2]{m}$ as
in the IDM. We have chosen this particular OV function in order to
reuse the 
steady-state IDM parameters, and also to compare the SFVDM with
Newell's car-following model~\cite{newell-carFollowing2002} on which the PCF model is based and which has a
tridiagonal fundamental diagram as well. Notice that, in the latter
model, $T=\tau=\unit[1]{s}$ is 
associated not only with the time gap $T$ but also with the reaction delay
time $\tau$~\cite{TreiberKesting-book}; furthermore, the  wave-speed
parameter of this model can be identified
by $w=(l\sub{veh}+s_0)/T$ where $l\sub{veh}$ is the vehicle length 
which, however, does not play a role in the platoon simulations). 

The top left panel of Fig.~\ref{fig:otherplatoon} shows an instance run of
the SFVDM simulations with the calibrated
dynamical parameters $1/\beta=\unit[10]{s}$,
$\lambda=\unit[0.52]{s^{-1}}$, and the white-noise intensity
$Q=\unit[0.25]{m^2/s^3}$. The top right panel displaying the speed standard
deviations along the platoon for 10 realisations can be compared with
that of the SIDM: Generally, we observe a good agreement. However, for
most realisations, the simulated
speed standard deviation of the first platoon vehicles is too high.
Simulating the FVDM with
action points (not shown) gives similar results as for the IDM. 

The middle row gives a typical run for the PCF model with the
recommended 
relaxation parameter
$1/\beta=\unit[16]{s}$, and the
calibrated noise intensity
$Q=\unit[1.05]{m^2/s^3}$ (only for the followers) while
the static parameters 
$v_0\sup{lead}=\unit[8.3]{m/s}$, $v_0=\unit[30]{m/s}$,
$T=\tau=\unit[1.0]{s}$, and $s_0=\unit[2]{m}$ are 
that of the other models. Notice that 
Ref.~\cite{laval2014parsimonious}
  recommends $1/\beta=\unit[16]{s}$ and a scaled noise intensity
  $\tilde{\sigma}^2=Q/(\beta v_0^2)=0.11^2$ (Section~4) and
  $\tilde{\sigma}^2=0.15^2$ (Section~5). With $v_0=\unit[30]{m/s}$,
  this gives $Q=\unit[0.68]{m^2/s^3}$ and $Q=\unit[1.27]{m^2/s^3}$, respectively,
  which are of the same order of 
  magnitude. (Notice that, for a given value of the scaled noise
  intensity, the physical noise intensity $Q$ depends strongly on the
  desired speed $v_0$ while only a weak dependency is plausible for the experiments.)
Both values of the noise intensity are significantly higher as that
calibrated for the stochastic IDM and FVDM.  This is caused by the
fact that the 
stochasticity of the PCF model is restricted to the free regime while, in the
interacting regime (the speed is restricted by the leader), the PCF
model reduces to Newell's model, i.e., 
the drivers follow deterministically the leader's trajectories with a constant space and time
shift and, consequently, have the same constant variance as that of
the leader. This applies whenever the realisation of the stochastic
free displacement is greater than the deterministic displacement
obtained from Newell's model since the minimum of the displacements is
taken in the model. The chance of deterministic car-following increases with
decreasing $Q$, increasing $v_0$, and increasing $\beta$. For
sufficiently low acceleration noise, this is always the case in our
experiment and the
PCF model reverts to deterministic car-following according to Newell's
model. This switching between a deterministic and a stochastic model
seems also to be the reason why, although only implementing the
acceleration noise mechanism,  the PCF model fits the observations
better than the SIDM and the SFVDM (and equally well as the IDM with
action points): For the first platoon vehicles, the leaders exhibit
only low-amplitude oscillations increasing the chance that the
deterministic part of the PCF model applies. This reduces the
effective acceleration noise of the first followers relative to that
of the vehicles further behind thereby reducing the
speed standard deviation of the first vehicles which is in line with
the observation. 

However, it is obviously unrealistic to switch from stochastic to
deterministic driving when entering the car-following regime. Therefore, a
straightforward generalization of the PCF model consists in adding the
same acceleration noise to
the car-following situation as well and also relax the rigid following
rule of Newell's model by introducing the same relaxation parameter
$\beta$ as for the free part. It can be shown that the
resulting ``Full-noise PCF model'' (FPCF model) is mathematically equivalent to~\refkl{micGen} where
$f$ is given by the time-delayed OVM with tridiagonal fundamental
diagram, 
\be
\label{FPCF}
f(s_n(t-T),v_n(t-T)) = \beta \left(v\sub{opt}(s(t-T))-v(t-T)\right),
\ee
with $v\sub{opt}(s)$ given by~{FVDM}. With the significantly reduced
noise intensity $Q=\unit[0.2]{m^2/s^3}$, this model reproduces the data
similarly well as the stochastic IDM and FVDM models which is to be
expected since it utilizes the noise mechanism while having no action
points and being at marginal stability (in Newell's model,
oscillations neither decay nor grow).

We conclude that the IDM is not necessary for our general
findings and can be replaced by other underlying car-following
models. In contrast, 
the mechanism matters.  While the instability mechanism cannot
reproduce the data even qualitatively, the action-point and noise
mechanisms reproduce the data nearly quantitatively with the
action-point mechanism and the selective-noise mechanism of the PCF
model giving marginally better results than unconditional acceleration
noise.

\section{\label{sec:bicycle}Bicycle Traffic on a Ring}
%
The purpose of this section is twofold: Firstly, we demonstrate that,
by a suitable change of the IDM parameters (particularly the vehicle
length and the desired speed), the 
IDM can also be used to realistically simulate bicycle
traffic. Secondly, we show that, for low-speed traffic, acceleration
noise alone can lead to fully developed stop-and-go waves even in the
subcritical regime.
 
\begin{figure}
\fig{0.99\textwidth}{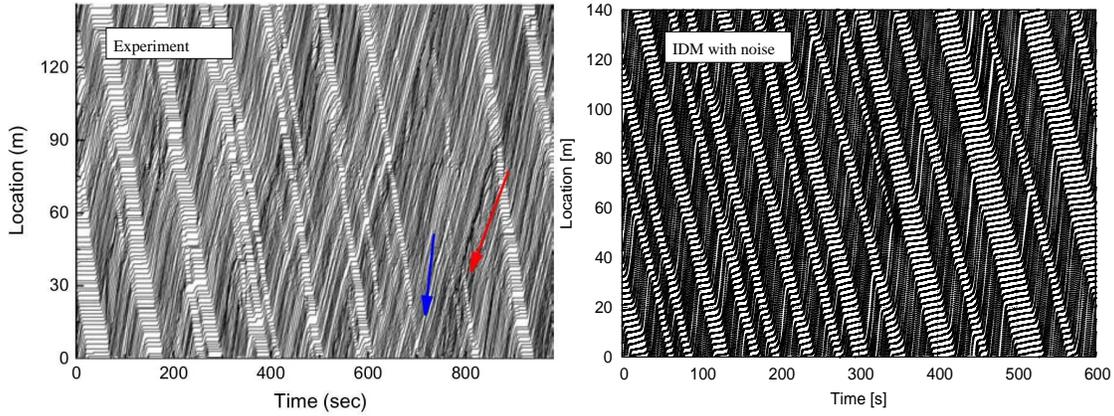}
\caption{\label{fig:bicycle}Left: experimental trajectories of bicycles on a closed
  ring course of \unit[140]{m} circumference; right: stochastic IDM
  simulation with
  acceleration noise 
  $Q=\unit[0.4]{m^2/s^3}$.  The IDM parameters corresponding to a
  subcritical situation are given in the main text.
}
\end{figure}

\begin{figure}
\fig{0.7\textwidth}{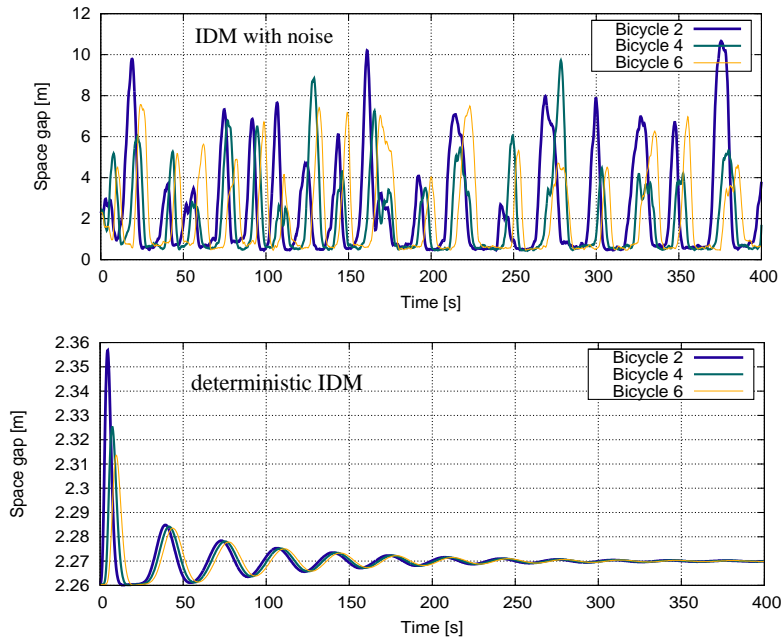}

\caption{\label{fig:bicycle-tseries}Top panel: gap time series of the
  simulation 
  of Fig.~\ref{fig:bicycle}; bottom panel: time series for the
  deterministic simulation with zero acceleration noise, $Q=0$. The
  remaining parameters are that of Figure~\ref{fig:bicycle}.
}
\end{figure}

Figure~\ref{fig:bicycle} shows simulations (right) of bicycle
experiments (left) on a ring
road of \unit[140]{m} circumference~\cite{jiang2016bicycle}. The
bicycle length is assumed to be \unit[1.67]{m} and the IDM
parameters have been set to $v_0=\unit[4]{m/s}$,
$T=\unit[0.6]{s}$, $s_0=\unit[0.4]{m}$, $a=\unit[0.8]{m/s^2}$, and
$b=\unit[1.5]{m/s^2}$. A
comparison of the simulated trajectories (right panel) with Fig.~3(c)
of Ref.~\cite{jiang2016bicycle} (left panel)  reveals a nearly quantitative agreement, at least in the
statistical sense. The nearly triangular and symmetrical fundamental
diagram (not shown) is consistent with the observations  as well.
Remarkably, the 
simulation is well in the subcritical regime ($\epsilon=-0.2$)
which is mainly caused by the low speeds. Nevertheless, the
acceleration noise alone
($Q=\unit[0.4]{m^2/s^3}$) leads to fully developed stop-and-go waves
which can also be seen by the gap time series of selected bikers 
(Fig.~\ref{fig:bicycle-tseries}, top panel). To validate this, we also
run simulations without noise and unchanged parameters,
otherwise (bottom panel). We observe that
the initial transients quickly dissipate which is consistent with
string stability.

\section{\label{sec:concl}Conclusion}
%
In this contribution, we have proposed a minimal general model for
simultaneously 
considering three possible mechanisms to traffic flow oscillations:
string instability, external white acceleration noise, and indifference
regions implemented by action points. Each of these mechanisms can be
activated and controlled independently from the others by varying a
single model parameter per mechanism, $\epsilon$, $Q$, and $\Delta
a\sub{max}$, 
respectively. The model is based on existing deterministic
time-continuous car-following models with a fundamental
diagram. However, the action-point mechanism introduces an
indifference region and \emph{de facto} converts this model into one
consistent with the three-phase theory of Kerner~\cite{Kerner2012physics}.

By analytical means  and numerical
simulations, we have shown that white acceleration noise as well as the
action points leads to highly spatiotemporally
correlated fluctuations of speeds and gaps that ``anticipate'' the
traffic waves produced by linear instabilities even well in the
linearly stable region ($\epsilon<0$). 

From the simulation results, we conclude that acceleration noise
and action points lead to similar results. This means that the
observed concave form of the fluctuation amplitude as a function of
the vehicle index in a platoon can not only be reproduced by models of
the three-phase theory, such as the 2D-IDM~\cite{tian2016improved}, but also by
``two-phase models'' such as the IDM when adding the simplest form of
stochasticity, white (uncorrelated) acceleration noise. 

We have also found that, for typical speeds of cars, acceleration
noise and action points alone will not lead to realistic traffic
oscillations. Additionally, at least marginal stability ($\epsilon=0$)
or a mild form of linear instability ($\epsilon$ slightly positive) is
necessary. In contrast, for low-speed traffic such as bicycle traffic,
acceleration noise (or action points) alone can lead to fully
developed and realistic traffic waves.

\bibliographystyle{elsart-num}
\bibliography{database}

\end{document}